\documentclass[a4paper,11pt]{article}
\usepackage[margin=2.5cm]{geometry}
\usepackage{setspace,graphicx,url,hyperref,enumitem}
\usepackage[linesnumbered,ruled]{algorithm2e}

\usepackage{amsmath,amssymb,amsfonts,amsthm,natbib}

\newcommand{\R}{\ensuremath{\mathbb{R}}}

\begin{document}
\onehalfspacing
\title{Reducing quantum annealing biases for solving the graph partitioning problem}
\author{Elijah Pelofske\footnote{Los Alamos National Laboratory, Los Alamos, NM 87545, USA},
Georg Hahn\footnote{Harvard University, T.H.\ Chan School of Public Health, Boston, MA 02115}, and Hristo N.\ Djidjev\footnotemark[1]}
\date{}
\maketitle

\begin{abstract}
    Quantum annealers offer an efficient way to compute high quality solutions of NP-hard problems when expressed in a QUBO (quadratic unconstrained binary optimization) or an Ising form. This is done by mapping a problem onto the physical qubits and couplers of the quantum chip, from which a solution is read after a process called quantum annealing. However, this process is subject to multiple sources of biases, including poor calibration, leakage between adjacent qubits, control biases, etc., which might negatively influence the quality of the annealing results. In this work, we aim at mitigating the effect of such biases for solving constrained optimization problems, by offering a two-step method, and apply it to Graph Partitioning. In the first step, we measure and reduce any biases that result from implementing the constraints of the problem. In the second, we add the objective function to the resulting bias-corrected implementation of the constraints, and send the problem to the quantum annealer. We apply this concept to Graph Partitioning, an important NP-hard problem, which asks to find a partition of the vertices of a graph that is balanced (the constraint) and minimizes the cut size (the objective). We first quantify the bias of the implementation of the constraint on the quantum annealer, that is, we require, in an unbiased implementation, that any two vertices have the same likelihood of being assigned to the same or to different parts of the partition. We then propose an iterative method to correct any such biases. We demonstrate that, after adding the objective, solving the resulting bias-corrected Ising problem on the quantum annealer results in a higher solution accuracy.
\end{abstract}

\section{Introduction}
\label{sec:introduction}
Quantum annealers of D-Wave Systems, Inc., offer a viable approach to compute high quality solutions of NP-hard problems that can be expressed as minimization of a quadratic form, i.e.,\ as minimization of a function of the form
\begin{align}
    H(x_1,\ldots,x_n) = \sum_{i=1}^n h_i x_i + \sum_{i<j} J_{ij} x_i x_j,
    \label{eq:hamiltonian}
\end{align}
where $h_i \in \R$ and $J_{ij} \in \R$ are specified by the user and define the problem under investigation. The variables $x_1,\ldots,x_n$ are discrete and unknown and need to be determined in order to minimize $H$. If $x_i \in \{0,1\}$ then eq.~\eqref{eq:hamiltonian} is called a QUBO (quadratic unconstrained binary optimization) problem, and if $x_i \in \{-1,+1\}$, then eq.~\eqref{eq:hamiltonian} is called an Ising problem. Both the QUBO and Ising formulations are equivalent \citep{Chapuis2019}. Many important NP-hard problems can be expressed as minimization of a function of the form of eq.~\eqref{eq:hamiltonian}, see \cite{Lucas2014}.

Problems expressed in QUBO or Ising form, that is via eq.~\eqref{eq:hamiltonian}, belong to the class of \textit{unconstrained optimization problems}. This means that the formulation in eq.~\eqref{eq:hamiltonian} contains both the objective function which is to be minimized, as well as a penalty term to enforce the constraints. Typically, this is realized by adding the penalty, scaled by some fixed prefactor, to the objective function. The prefactor of the penalty is chosen large enough such that it is never favorable to break the constraints in order to achieve a further reduction in the objective. An example of an unconstrained optimization in the form of eq.~\eqref{eq:hamiltonian}, with the scaled constraint added to the objective, is given in Section~\ref{sec:biases}.

In order to obtain an approximate solution of the minimization of eq.~\eqref{eq:hamiltonian} with the D-Wave 2000Q device, the problem under investigation has to be mapped onto the annealer so that linear coefficients $h_i$ are mapped onto qubits and quadratic coefficients $J_{ij}$ are mapped onto \textit{couplers} connecting pairs of qubits. After a process  called \textit{quantum annealing}, a solution is read off the annealer. This operation is subject to a variety of sources of bias, which might negatively influence the annealing accuracy, for instance:
\begin{enumerate}
    \item Although the coefficients $h_i$ and $J_{ij}$ in eq.~\eqref{eq:hamiltonian} can be chosen arbitrarily, they need to be rescaled to $h_i \in [-2,2]$ and $J_{ij} \in [-1,1]$ and rounded to 8 bits in order to be mapped as currents onto the chip by a digital-to-analog converter. Both the resolution reduction to 8 bits and the conversion potentially introduce biases.
    \item Variations in manufacturing can lead to some qubits behaving slightly differently than others.
    \item The physical qubits on the D-Wave quantum chip are arranged in a particular graph structure, called \textit{Chimera} graph \citep{LANLDwave,Chapuis2019}. However, the structure of the non-zero couplers in eq.~\eqref{eq:hamiltonian} does not necessarily match the Chimera architecture, thus requiring a \textit{minor embedding} of the QUBO or Ising connectivity onto the Chimera graph. Moreover, in the embedding, several physical qubits are often identified to act as one logical qubit, called a \textit{chain}. Since the embedding and thus the chains stay fixed during annealing, their choice potentially impacts the solution quality.
    \item So called \textit{leakage} on the physical chip from a coupler connecting to its incident qubits can modify their intended behavior, thus changing the effective value of the weight assigned to them \cite{dwave_gauge}. This effect is reported to be more severe for chained qubits.
\end{enumerate}
As a result, various works in the literature have already showed that the D-Wave annealers are biased samplers \citep{Mandra2017,King2016,Albash2015,Boixo2013}.

In this contribution, we try to mitigate the effect of such biases with the aim to improve annealing results of constrained minimization problems, and apply our concept to Graph Partitioning, an important NP-hard problem. Briefly, let $G=(V,E)$ be a graph consisting of a vertex set $V=\{1,\ldots,n\}$ and an edge set $E \subseteq V \times V$. The Graph Partitioning problem asks us to divide the set of vertices $V$ into two disjoint and balanced sets $V_1$ and $V_2$, satisfying $V = V_1 \cup V_2$ and $V_1 \cap V_2 = \emptyset$, such that the size of $V_1$ and $V_2$ differs by at most one (the balance constraint) and the number of \textit{cut edges} $\{ e=(v,w): v \in V_1, w \in V_2 \}$ between the two partitions is minimized (the objective). An Ising formulation of the Graph Partitioning problem, given in \cite{Lucas2014}, is considered in Section~\ref{sec:biases}.

We attempt a reduction of hardware biases with the help of a two-step method. In the first step, we aim at quantifying the biases that result from implementing the constraint of the problem. In the case of Graph Partitioning, these are the biases resulting from mapping the balance constraint onto the annealer. To be precise, if no objective function (the number of cut edges between the two partitions) is given, we would expect any assignment of vertices satisfying the balance constraint to be returned equally likely from the annealer. However, we will show that this is not the case. We propose an algorithm to iteratively modify the coefficients of the constraint only, with the aim to reduce the measured biases. Once the biases of the constraint are corrected, in the second step, we add the objective function to the corrected constraint and send the new formulation to the D-Wave annealer.

Having an unbiased implementation of a constraint is an important factor in getting optimal or high-quality solutions of constrained optimization problems. If the implementation is biased towards some combinations of variables, then such samples can be preferred even if the value of the objective function for them is suboptimal. On the other hand, a  feasible solution with optimal value of the objective may be neglected by the annealer if the implementation of its constraint is biased against it.

Using Erd{\H o}s–R{\'e}nyi graphs as an example, our work shows that for Graph Partitioning the mapped constraint is considerably biased, and that our iterative algorithm is able to correct those biases. After adding the objective, we show that submitting the corrected problem to the D-Wave annealer results in more accurate solutions after annealing than obtained with the original formulation, measured in terms of a lower cut value of the partitioning (while satisfying the balance constraint).

This article is structured as follows. After a detailed literature review in Section~\ref{sec:literature_review}, we introduce a method to measure biases occurring in the constraint of the Graph Partitioning problem as well as an algorithm to correct them (Section~\ref{sec:methods}). In Section~\ref{sec:experiments}, we apply our methodology to random instances of the Graph Partitioning problem, and show that mapping the problem onto the D-Wave quantum chip indeed incurs biases. We demonstrate that after correcting the constraint, those biases can be significantly reduced. Finally, we show that solving the corrected formulation (with added objective) instead of the original Ising problem yields more accurate solutions. The article concludes with a discussion in Section~\ref{sec:discussion}.

\subsection{Literature review}
\label{sec:literature_review}
Quantifying and correcting sources of biases/errors of a quantum annealer is an active area of research. Several approaches pursued in the literature are worth mentioning. First, some works aim for physical shortcomings of a device. For instance, \cite{Pudenz2014} attempt to correct for decoherence in quantum annealing, which causes quantum superpositions to decay into mutually exclusive classical alternatives, thereby losing accuracy. According to \cite{Pearson2019}, analog control noise can make the probability that the implemented Hamiltonian shares a ground state with the intended Hamiltonian exponentially small in the problem size and the magnitude of the noise. The authors counteract this by showing empirically that the simple method of \cite{Pudenz2014} to correct errors brings down the TTS (time-to-solution) to below the one of classical solvers.

Second, some approaches address the problem of error correction of a quantum device by viewing it essentially as a decoding problem. For instance, the problem of interpreting the output of a quantum error correction code is considered in \citep{Roffe2019}, known as "decoding". In their paper, the authors consider the use of a quantum annealer for decoding, and they introduce a quantum error correction protocol based on the coherent parity check (CPC) framework for quantum error correction.
Two heuristics, called "single qubit heuristic" and "multi qubit correction" are introduced in \cite{Ayanzadeh2020} with the aim to virtual tunnels, that is, pairs of qubits which, if changed simultaneously, can result in a new state of lower energy.
To guard against single-qubit errors, \cite{Jiang2017} proposes to use a Hamiltonian consisting of sums of the gauge generators from so-called Bacon–Shor codes for error correction.
In \cite{Li2020}, the authors address the increased effective temperature of a physical quantum device, which will negative influence the distributions from which a quantum annealer draws samples. The authors aim to address this by error correction schemes which can reduce the effective temperature, which is achieved by mapping the input Hamiltonian to an error correcting "repetition code" Hamiltonian, which in turn is then mapped onto the physical hardware qubits.

Third, another strategy to correct errors relies on the introduction of energy penalties \citep{Bookatz2015}. Here, a constant (time-independent) term is added to the Hamiltonian which penalizes states that have been corrupted by e.g.\ single-qubit errors, in the hope that the penalty will impose an energy barrier that must be surpassed for an error to occur.
In \cite{Perdomo2016}, the authors attempt a similar objective as we do, that is "to determine and correct persistent, systematic biases between the actual values of the programmable parameters and their user-specified values". However, they follow a different approach, in which a thermal model for only the linear or quadratic contributions in eq.~\eqref{eq:hamiltonian} is assumed, and measurements from the D-Wave annealer are fitted to the model in order to quantify their biases.

Finally, reducing the hardware bias has been realized by carefully choosing appropriate annealing control parameters. Reverse annealing \cite{Chancellor2017, Ohkuwa2018} was proposed to change the standard annealing schedule so that to allow a local search in a neighborhood of a previously known suboptimal solution. Alternative schedules for local search based on the h-gain feature of D-Wave were proposed in \cite{hgain-QCE}. Optimizing annealing parameters linked to a fixed embedding was investigated in \cite{barbosa2020optimizing} and shown to improve the accuracy of the solutions.

On a related note, in \cite{Koenz2019} the authors provide indications that complex driver Hamiltonians beyond transverse fields are not able to mitigate sampling biases, thus suggesting that quantum annealing machines are not well suited for sampling applications unless postprocessing techniques to improve the sampling are applied, see for example \cite{pelofske2020advanced}.

\section{Methods}
\label{sec:methods}
In this section we describe our approach. First, we detail our method for quantifying the biases occurring in the implementation of the constraint of the graph partitioning problem (Section~\ref{sec:biases}). Second, we propose an iterative algorithm for reducing those biases by modifying the coefficients of the  Ising problem representing it (Section~\ref{sec:algorithm}). Third, after having corrected the constraint, we assemble a new Ising model by adding the objective function to the modified constraint (Section~\ref{sec:combined_ising}).

\subsection{Quantifying biases}
\label{sec:biases}

Since a constraint in an optimization problem is implemented in an Ising model as a penalty function with energy (value) zero if the assignments of the variables (samples) is feasible, or positive energy if the samples are infeasible, all feasible samples have the same constraint energy. Therefore, in the absence of an objective function, all feasible samples should be uniformly represented in an unbiased sampler. If the sampler is biased, we can measure the bias of the samples with the help of an appropriate (problem-specific) metric, which is then used to debias the constraint.

The following illustrates this idea in the case of Graph Partitioning (GP). 
In the Ising formulation of GP \cite{Lucas2014}, a variable $x_i\in\{-1,1\}$ is assigned to each vertex $i$, and depending on whether $x_i$ takes value $+1$ or $-1$, vertex $i$ is assigned to either the "+" or the "--" partition. With no objective present, samples from the annealer satisfying the constraint should have the property that any pair of logical variables $(x_i,x_j)$, where $i,j \in \{1,\ldots,n\}$ and $i<j$, should be allocated equally likely to either the same or to different partitions. Therefore, if we denote by $n_{ij}$ how many times $x_i$ and $x_j$ share the same partition (i.e., $x_i=x_j$) among $N$ anneals, the fraction $\frac{n_{ij}}{N}$ should approach $0.5$ for all variable pairs as $N \rightarrow \infty$. This holds true if the sampling was indeed unbiased. We call $b_{ij} = \frac{n_{ij}}{N}-0.5$ for $i,j \in \{1,\ldots,n\}$, $i<j$, the \textit{quadratic bias}.

For a given set of couplers $J = \{J_{ij}: i < j\}$, we will denote with \textit{calculate\_bias(J)} the function computing the set of all $b_{ij}$, $i<j$, according to the definition above.

According to \cite{Lucas2014}, the Ising formulation for GP (with two partitions) is given as
\begin{align}
    H(x_1,\ldots,x_n) = A \left( \sum_{i=1}^n x_i \right)^2 + B \sum_{(i,j)\in E} \frac{1-x_i x_j}{2},
    \label{eq:gp}
\end{align}
where $G=(V,E)$ is the graph to be partitioned, $n = |V|$, $x_i \in \{-1,+1\}$ are the spin indicators which specify if vertex $i$ is allocated to either the "+" or "--" partition. The first term with prefactor $A$ is the balance constraint, whereas the second one with prefactor $B$ is the objective, which is the number of cut edges between the two partitions that we aim to minimize. According to \cite{Lucas2014}, any choice of $A$ and $B$ satisfying $A/B \geq n/8$ ensures that it is never favorable to break the balance constraint in order to achieve a further reduction in the number of cut edges. Without loss of generality, we can choose $B=1$ and $A=n/8$. Since the largest graph with arbitrary connectivity that can be embedded onto D-Wave 2000Q contains $n=65$ vertices (variables), we fix $A=9$ and $B=1$ in the remainder of the article to guarantee that $A/B \geq n/8$. To measure the quadratic bias, we consider the quadratic constraint $\left( \sum_{i=1}^n x_i \right)^2$ of eq.~\eqref{eq:gp} only.

\subsection{Bias correction procedure}
\label{sec:algorithm}
After having computed the bias $b_{ij}$ for all $i<j$, we modify the quadratic terms $\sum_{i<j} J_{ij} x_i x_j$ of the GP constraint implementation in eq.~\eqref{eq:gp}, where $J_{ij}=2A$ (multiplying out the constraint in eq.~\eqref{eq:gp} shows that each quadratic coupler is $2A$), as follows.

Obviously, larger biases should require more correction. Therefore, the magnitude with which the quadratic term $J_{ij}$ in the constraint (coupling variables $i$ and $j$) is modified should depend on the bias between the variables $x_i$ and $x_j$. Furthermore, if $b_{ij}>0$ for a pair $(i,j)$, the two vertices $i$ and $j$ evidently appear too often in the same partition, which is due to their weight $J_{ij}$ being too small. This is because terms of the form $J_{ij} x_i x_j$ with large $J_{ij}>0$ increase the objective value if $x_i=x_j$, thus making an assignment $x_i=x_j$ less favorable. Therefore, for each variable pair $(i,j)$, we use the update rule
\begin{align}
    J_{ij}' = J_{ij}+kb_{ij},
    \label{eq:update}
\end{align}
where $b_{ij}$ is the empirical bias for the pair $(i,j)$, $J_{ij}$ is the original quadratic weight, $J_{ij}'$ is the bias-corrected weight, and $k>0$ is a scaling constant to be determined.

Note that, as with all quantum computers, D-Wave is a probabilistic sampler, meaning that we would expect deviations of $b_{ij}$ from $0$ even if the annealer were unbiased. It is thus sensible to only correct a bias once it surpasses a given threshold $\tau>0$ (that is, only when $|b_{ij}|>\tau$). The parameter $\tau$ is likewise chosen in advance.

With the function \textit{update\_terms} we will denote the application of the aforementioned correction to all terms $J_{ij}$, $i<j$, of an input quadratic model $J = \{J_{ij}: i < j\}$, using the biases $b=\{b_{ij}: i < j\}$ and the parameters $k$ and $\tau$. The function returns a set of updated couplers $J' = \{J'_{ij}: i < j\}$ for all $i<j$ computed according to eq.~\eqref{eq:update}.

\begin{algorithm}[t]
    \caption{Iterative quadratic bias reduction\label{algo:quadratic_bias_reduction}}
    \SetKwInOut{Input}{input}
    \SetKwInOut{Output}{output}
    \Input{\ Ising quadratic terms $J$, scaling constant $k>0$, noise cutoff $\tau>0$, stopping threshold $\sigma>0$\;}
    \Output{\ Quadratic terms $J$ of optimized Ising\;}
    $b \leftarrow$ calculate\_bias$(J)$\;
    \While{$\min(b) < -\sigma$ or $\max(b) > \sigma$}{
        $J \leftarrow$ update\_terms$(J, b, \tau, k)$\;
        $b \leftarrow$ calculate\_bias$(J)$\;
    }
    \Return{$J$}\;
\end{algorithm}
A pseudo-code of our method is given as Algorithm~\ref{algo:quadratic_bias_reduction}. Its input are the quadratic terms $J = \{J_{ij}: i < j\}$ to be corrected, the scaling constant $k>0$ in eq.~\eqref{eq:update}, the aforementioned noise cutoff $\tau>0$, and some stopping threshold $\sigma>0$.

After calculating the biases with the help of the function \textit{calculate\_bias}, we iteratively correct the quadratic terms using the function \textit{update\_terms} and recalculate the biases with the function \textit{calculate\_bias}. The algorithm stops if the all biases are less than $\sigma$ in absolute value, for some stopping threshold $\sigma>0$ chosen in advance. This is to avoid an infinite loop in case the updated couplers diverge. The smaller $\sigma$, the more the algorithm will try to correct biases, and the longer it will run. While setting $\sigma=\tau$ would work in most cases, our experience shows that choosing $\sigma$ slightly larger than $\tau$, e.g., $\sigma=0.2$ and $\tau = 0.15$, leads to a better efficiency/accuracy tradeoff, and we therefore keep these two parameters separate in Algorithm~\ref{algo:quadratic_bias_reduction}.

Algorithm~\ref{algo:quadratic_bias_reduction} works directly on the logical variables, meaning it modifies the couplers of all chained qubits simultaneously.

\subsection{Adding objective information to the corrected Ising}
\label{sec:combined_ising}
After having run Algorithm~\ref{algo:quadratic_bias_reduction} to completion, the resulting couplers, which we denote as $J' = \{J_{ij}': i < j\}$, have the property that the associated Ising model (which contains only quadratic terms) yields unbiased samples, in the sense that any pair of variables in those samples is roughly equally likely to appear in either the same or opposing partitions.

To arrive at a corrected Ising model for GP, we only need to add the objective function information of GP to the modified constraint. To this end, we first normalize the corrected couplers $J'$ by dividing each coefficient with the maximum coefficient, and call the resulting set $J''$. This yields the bias-corrected constraint
$$C(x_1,\ldots,x_n) = \sum_{i<j} J_{ij}'' x_i x_j.$$
We then multiply the normalized constraint by $2A$ again (see Section~\ref{sec:biases}) and substitute it in lieu of the original constraint $\left( \sum_{i=1}^n x_i \right)^2$ in the Ising model of eq.~\eqref{eq:gp}. This yields the bias-corrected Ising model of GP as
\begin{align}
    H'(x_1,\ldots,x_n) = 2A \cdot C(x_1,\ldots,x_n) + B \sum_{(i,j)\in E} \frac{1-x_i x_j}{2}.
    \label{eq:gp2}
\end{align}
We investigate $H'$ in the experiments of Section~\ref{sec:experiments}.

\section{Experiments}
\label{sec:experiments}
In this section, we apply Algorithm~\ref{algo:quadratic_bias_reduction} to correct the constraint of GP as outlined in Section~\ref{sec:algorithm}, showing that indeed there is a considerable initial bias, which we are able to correct (Section~\ref{sec:training}). Importantly, we demonstrate that the new formulation $H'$ of GP in eq.~\eqref{eq:gp2} allows one to find better partitions (in the sense that they have a smaller edge cut while still satisfying the balance constraint) compared to the original Ising formulation of eq.~\eqref{eq:gp}.

In the following, we always solve GP on Erd{\H o}s–R{\'e}nyi random graphs with $|V|=65$ vertices and an edge probability $p$ uniformly sampled in $[0.05,0.95]$. We use $100$ anneals when running Algorithm~\ref{algo:quadratic_bias_reduction} to debias the constraint, and evaluate the final Ising model that includes the objective using $10000$ anneals. The annealing time is set to $1$ microsecond on D-Wave 2000Q. The chain strength is always determined with the help of D-Wave's \textit{unform torque compensation} feature, see \cite{dwave_torque}, using a prefactor of $1.5$. However, our experiments show that the constraint also exhibits a bias for longer annealing times, and that Algorithm~\ref{algo:quadratic_bias_reduction} is also able to correct that bias for longer annealing times.

\subsection{Debiasing the constraint of Graph Partitioning}
\label{sec:training}
We assume we are given an arbitrary graph $G=(V,E)$ of $|V|=65$ vertices, the maximum size of a fully connected graph embeddable on the D-Wave 2000Q.
To run Algorithm~\ref{algo:quadratic_bias_reduction}, we use the following set of parameters. We fix $k=10$ to correct the quadratic couplers according to eq.~\eqref{eq:update}, a choice which turned out to be advantageous in our simulations. A scheme in which $k$ varies over time is equally possible and remains a topic for further investigation. The bias threshold is set empirically at $\tau=0.05$. Clearly, choosing $\tau$ too small will lead to an overcorrection of biases that actually result from random fluctuations, whereas too large values of $\tau$ will result in biases not being corrected, although they are large enough to affect the annealing outcome. Similarly, regarding the stopping parameter $\sigma$, if $\sigma$ is too small, Algorithm~\ref{algo:quadratic_bias_reduction} might never terminate, while if $\sigma$ is too large, it will result in Algorithm~\ref{algo:quadratic_bias_reduction} finishing faster, but at the expense of not correcting all singnificant biases. We set $\sigma=0.2$ for the experiments in this article.

\begin{figure*}
    \centering
    \includegraphics[width=0.45\textwidth]{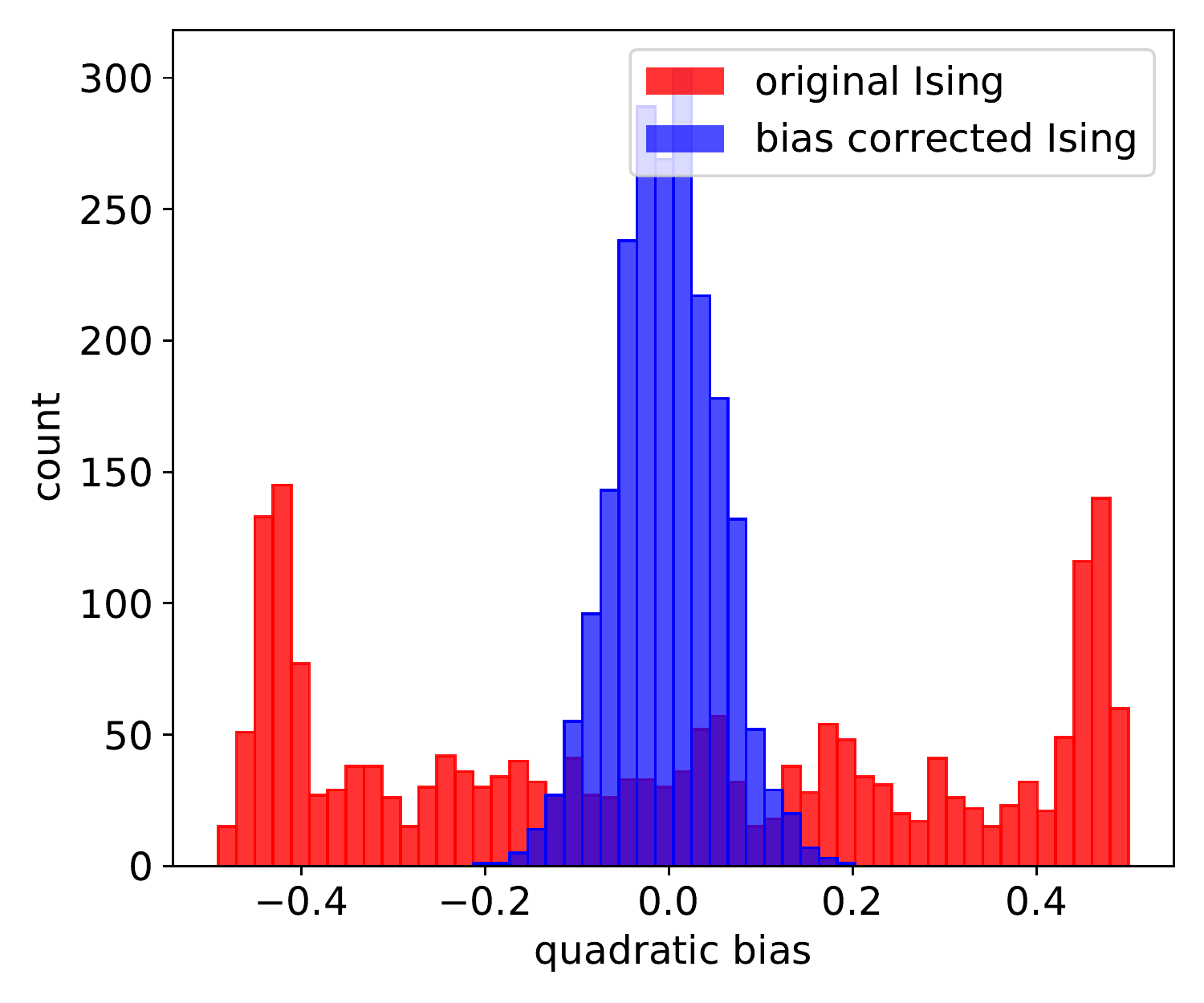}\hfill
    \includegraphics[width=0.45\textwidth]{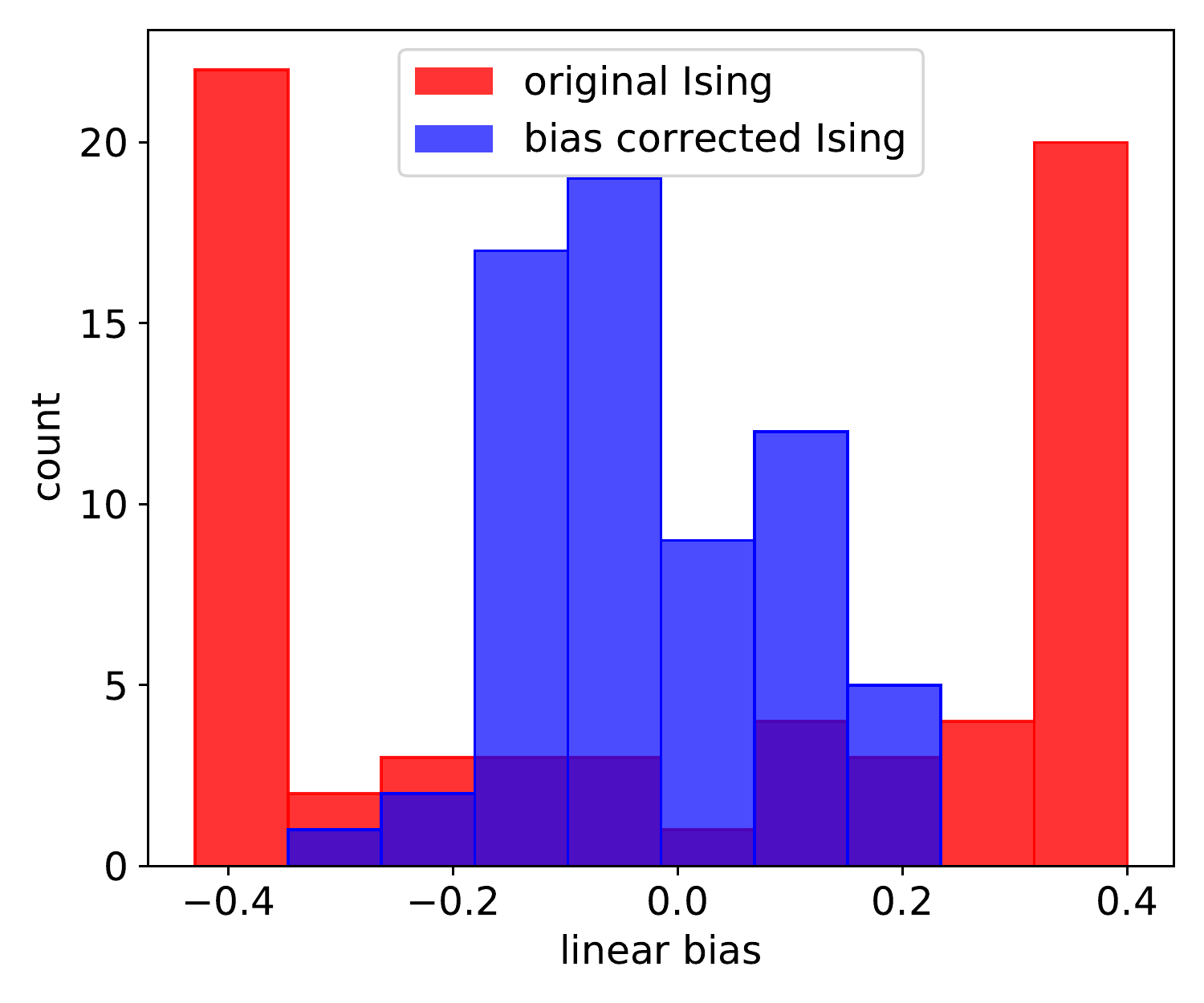}
    \caption{Left: distribution of quadratic biases at the first (red) and after the last (blue) iteration of Algorithm~\ref{algo:quadratic_bias_reduction} (total of $224$ iterations). Right: distribution of linear biases at the first (red) and after the last (blue) iteration in the same run of Algorithm~\ref{algo:quadratic_bias_reduction}.}
    \label{fig:histogram}
\end{figure*}
Results for biases before and after their correction are shown in Figure~\ref{fig:histogram} based on a single run of Algorithm~\ref{algo:quadratic_bias_reduction}. We focus first on the quadratic biases (left plot), containing $65 \cdot 64/2 = 2080$ datapoints (one for each pair of vertices). Several observations are noteworthy. First, after implementing the constraint of GP onto the D-Wave 2000Q, it is indeed not the case that any two random vertices share a partition with probability $0.5$, even though we would expect such a behavior in the absence of an objective (red histogram). Instead, we observe that the bias distribution is bimodal and varies from $-0.5$ to $0.5$, meaning many pairs of vertices are consistently in the same partition or consistently in different partitions even without having an objective term in the Ising model. After running Algorithm~\ref{algo:quadratic_bias_reduction}, we see that the quadratic biases are mostly corrected, yielding a new Ising model having the property that most pairs of variables indeed share a partition with probability $0.5$ and, equivalently, a bias of zero (blue histogram).

Though we do not correct linear biases, i.e., the coefficients $h_i$ in eq.~\eqref{eq:hamiltonian}, we can define a \textit{linear balance bias measure} $b_i$ as follows. For a given vertex $i \in \{1,\ldots,n\}$, we count how many times it comes out as $+1$ and how many times it is $-1$ after annealing when considering the constraint only. Letting $m_i$ be the count of $+1$ states among $N$ anneals for vertex $i$, we define $b_i = m_i/N-0.5$. This metric should approach $0$ for all variables if the assignment of each vertex to one of the two partitions is truly random.

Figure~\ref{fig:histogram} (right) shows the linear bias among all $65$ variables. We observe a similar picture as the one of the quadratic biases, even though we do not directly correct linear biases in Algorithm~\ref{algo:quadratic_bias_reduction}. Before the correction, most linear biases seem to be either -0.5 or 0.5, meaning that each of the corresponding variables is assigned the same value in almost all samples, rather than the values $+1$ and $-1$ with roughly the same frequency.
After the quadratic bias correction, most linear biases seem to get fixed as well as they cluster around the zero.

Though not displayed, we observe more outliers to the right for larger values of $\sigma$ (the stopping threshold for Algorithm~\ref{algo:quadratic_bias_reduction}).

\begin{figure*}
    \centering
    \includegraphics[width=0.45\textwidth]{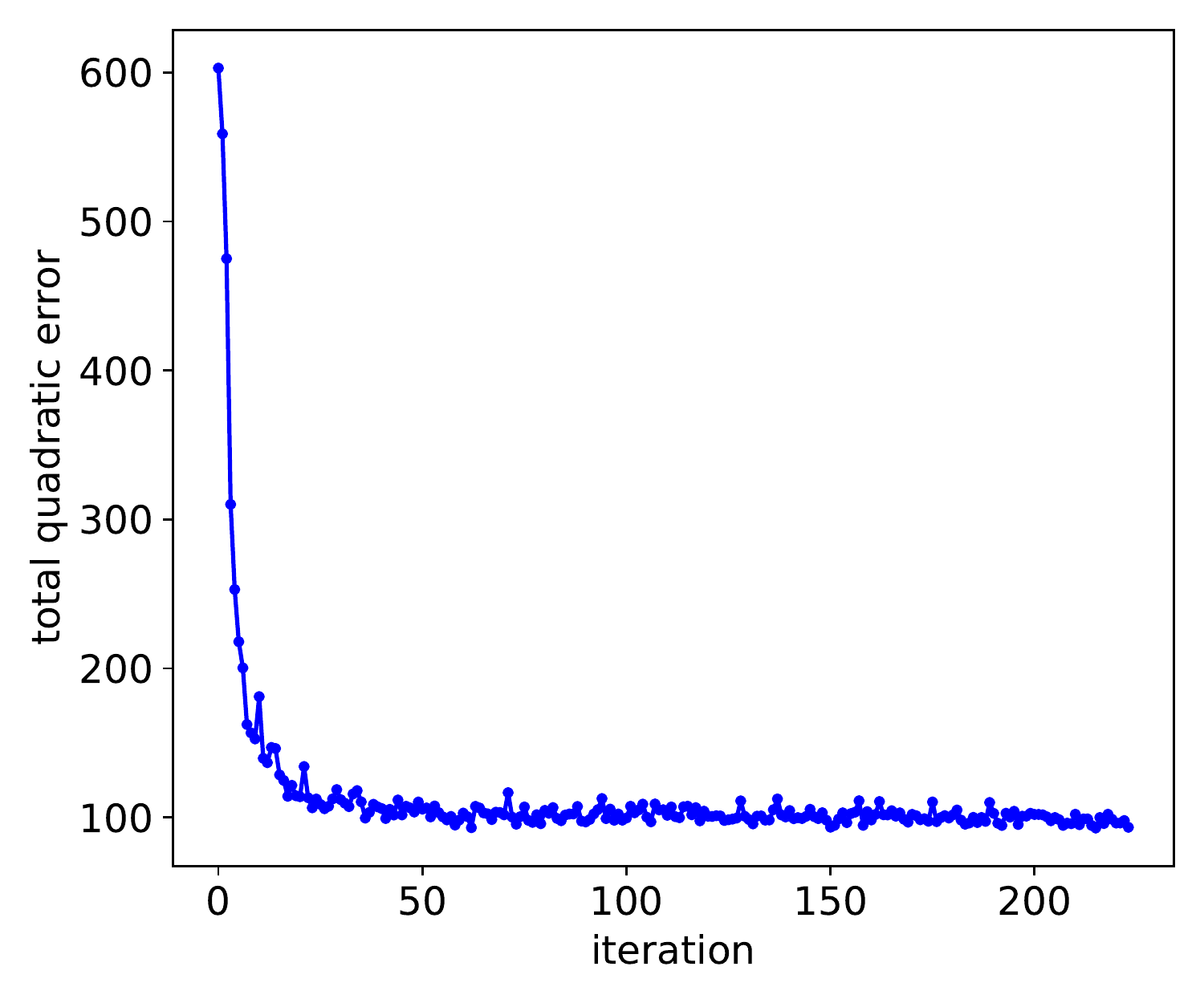}\hfill
    \includegraphics[width=0.45\textwidth]{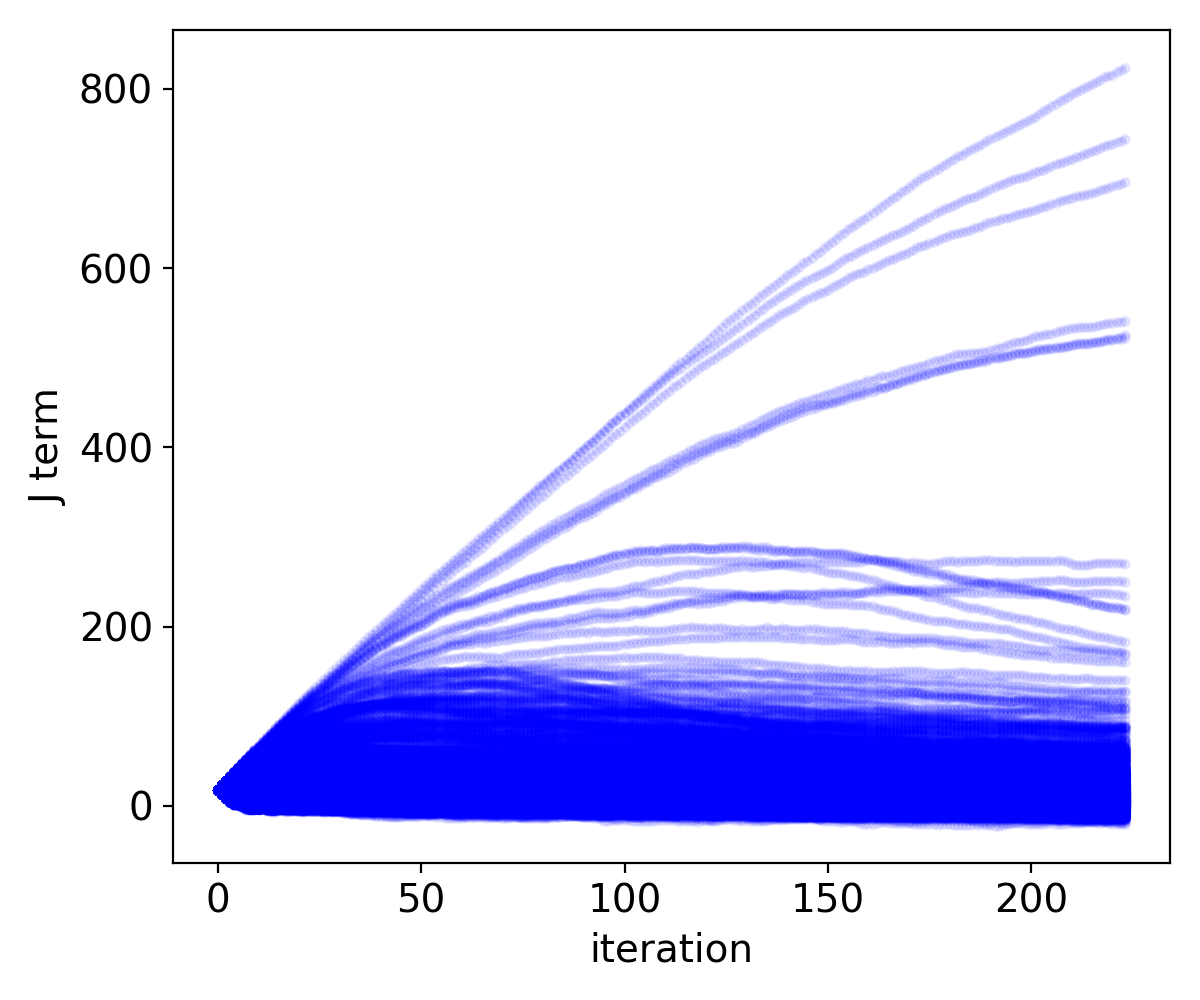}
    \caption{Left: total quadratic bias as a function of the iteration number. Right: the magnitude of the $J$ coefficients in the bias-corrected Ising model as a function of the iteration number. Each line corresponds to one  coefficient.}
    \label{fig:error_iteration}
\end{figure*}
For the same experiment  depicted in Figure~\ref{fig:histogram}, Figure~\ref{fig:error_iteration} shows the progression of the iterations of Algorithm 1, plotting the sum of all quadratic biases as a function of the iteration number (left). We observe that, as expected, the quadratic bias quickly decreases (within the first $50$ iterations), and stabilizes afterwards. Since the measurements from D-Wave 2000Q are noisy, we do not expect the quadratic bias to go to zero, and indeed it stabilizes at around $100$ (for the $2080$ pairs of variables, meaning an average bias of around $0.05$ per pair).

Figure~\ref{fig:error_iteration} (right) shows the progression of the individual quadratic couplers as Algorithm~\ref{algo:quadratic_bias_reduction} progresses. We observe that many quadratic couplers remain roughly unchanged (those in the range $[0,160]$). However, it seems that a handful of quadratic couplers cannot be corrected (these couplers have weights in the range $[400, 800]$ after bias correction). Those are the ones which consistently appear in the same partition, in which case Algorithm~\ref{algo:quadratic_bias_reduction} will increase their values to penalize such an assignment (see Section~\ref{sec:algorithm}). Apparently, it is not possible to fully correct their biases in this way. It is notable that this bias is asymmetrical, i.e., we do not see variables which also consistently do not appear in the same partition. The precise reason of this phenomenon is unknown, but could be related to their proximity on the D-Wave chip, or other hardware specifics.

\subsection{Solving a Graph Partitioning problem after the bias correction}
\label{sec:validation}
After having corrected the constraint of GP with the help of Algorithm~\ref{algo:quadratic_bias_reduction}, we proceed by adding back in the objective function (the edge cut), see Section~\ref{sec:combined_ising}. The following experiments are performed using the debiased Ising model $H'$ of eq.~\eqref{eq:gp2}, using the debiased constraint $C$ we calculated in the previous section.

For the experiments reported in this section, we use the same annealing parameters as in the previous section, with the only exception being that we now use $10000$ anneals for each D-Wave call.

\begin{figure}
    \centering
    \includegraphics[width=0.45\textwidth]{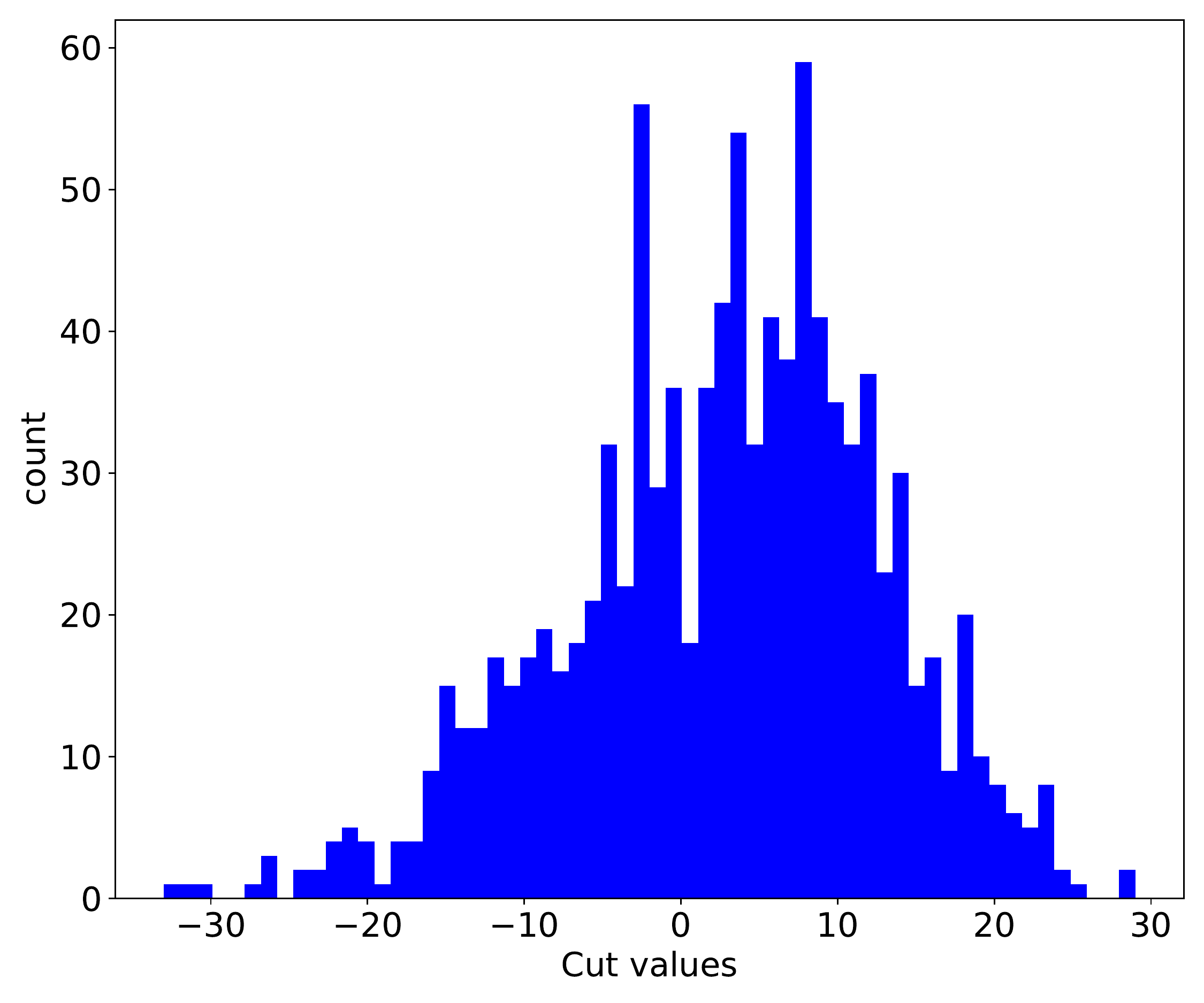}
    \caption{Difference in cut value of the original Ising model for GP and the bias-corrected one from Section~\ref{sec:combined_ising}. Since the objective is to minimize the edge cut, a positive mean of the distribution implies that the average cut value of the bias-corrected Ising model of Algorithm~\ref{algo:quadratic_bias_reduction} is smaller than the one of the original formulation.}
    \label{fig:validate_histogram}
\end{figure}
We assess whether the bias-corrected Ising model $H'$ for GP produces better solutions compared to the original one $H$. We generate 1000 random instances of GP and solve them on D-Wave 2000Q using both Ising formulations. For each instance, we find the best cut size found using each of the two formulations, and record their difference. We discard any sample that does not satisfy the balance constraint, since otherwise a comparison in terms of the cut size is not sensible. Hence, we end up with 1000 data points, one per instance, each showing the difference between the best cut size found using the original formulation, and the one found using the bias-corrected one. Since the objective of GP is to minimize the edge cut, a positive difference implies that the average cut value of the bias-corrected Ising model of Algorithm~\ref{algo:quadratic_bias_reduction} is smaller than the one of the original formulation.

Figure~\ref{fig:validate_histogram} shows the results. We observe that the histogram of edge cut differences is indeed shifted to the right, thus demonstrating that the bias-corrected Ising model for Graph Partitioning does yield lower (i.e., better) edge cuts. To be precise, the mean of the histogram depicted in Figure~\ref{fig:histogram} is $2.8$ (the improvement of our bias-corrected Ising over the original formulation in eq.~\eqref{eq:gp} in the average case), out of a mean edge cut of $496$ for the original formulation alone. Out of the $1000$ random problem instances we submitted to D-Wave, $621$ had an improved cut value with the bias-corrected Ising (compared to the original Ising model), and $343$ had a worse cut value.

\section{Discussion}
\label{sec:discussion}
In this work, we show that samples obtained with D-Wave 2000Q are biased, even in the absence of the objective function of the Ising model being solved. We attempt to quantify the biases occurring when implementing the constraint and propose an algorithm to modify the quadratic couplers in order to arrive at an unbiased version of it.

Adding back the objective function of the problem instance under consideration, that is, the edge cut in the case of GP considered in this article, allows us to arrive at a bias-corrected Ising model. We empirically show that, indeed, the bias-corrected constraint performs more in line with what would be expected for a random assignment of $+1$ and $-1$ to all variables. Importantly, we demonstrate that the bias-corrected Ising formulation for GP allows us to obtain samples of higher quality, meaning balanced partitionings with a lower edge cut, than the ones obtained by solving the original formulation of GP on the D-Wave 2000Q.

An advantage of our approach is that, in the case of GP, the constraint does not depend on the structure of the input graph, but only on the number of its vertices. That means that, once we find an unbiased implementation of the constraint for some graph of $n$ vertices, we can directly use the same implementation of the constraint for solving GP on any $n$-vertex graph.

This work is a first step towards iterative bias correction and leaves scope for several avenues of future work:
\begin{enumerate}
    \item As a first step, we only correct quadratic biases in this contribution, as this turned out to be most effective in our experiments. Extending our approach to linear bias corrections remains for future work.
    \item Though our approach is more general, we only present results for GP in this article. We plan to consider more problems beyond GP in subsequent works.
    \item Further work is necessary to understand why some variables are more biased than others. Likewise, the role of the fixed embedding we use deserves a closer look.
    \item The selection of the tuning parameters $k$, $\sigma$, and $\tau$ of Algorithm~\ref{algo:quadratic_bias_reduction} was done in a rather ad-hoc fashion. Optimizing them with the help of, e.g., a genetic optimization algorithm, could lead to both a more effective bias reduction and better annealing solutions of the bias-corrected Ising problems.
    \item For GP, experimental observations suggest that the bound $A/B \geq n/8$ stated in Section~\ref{sec:biases} can be relaxed. This would allow us to have less extreme weights in the Ising model, which typically results in higher quality samples returned by D-Wave 2000Q. The optimal choice of $A$ and $B$ remains to be investigated.
    \item There are other ways to manipulate the bias of a chained problem on D-Wave which remain to be explored. In particular, methods such as annealing offsets, spin reversals, and varying how the weights across chains are distributed, are all potential alternative methods.
\end{enumerate}


\end{document}